\documentclass[twoside,fleqn]{article}
\usepackage{espcrc2,latexsym}
\usepackage{amsmath}
\usepackage{graphicx}
\usepackage{amsfonts}
\usepackage{amssymb}
\usepackage{epsfig}

\title{Zone methods and the fermion sign problem}
\author{Dean Lee\thanks{The author thanks Shailesh Chandrasekharan, Hans Evertz, Pieter Maris, and Nathan Salwen for discussions and comments.  This work benefitted from the resources of the North 
Carolina Supercomputer Center.}
\address{Dept. of Physics, North Carolina State University, Raleigh, NC 27695}}

\begin{document}

\begin{abstract}
We review a recently proposed approach to the problem of alternating 
signs for fermionic
many body Monte Carlo simulations in finite temperature simulations.
We derive an estimate for fermion wandering lengths and 
introduce the notion of permutation zones, special regions of 
the lattice where identical fermions may interchange and outside 
of which they may not.  Using successively larger permutation zones, 
one can extrapolate to obtain thermodynamic observables in regimes 
where direct simulation is impossible.
\vspace*{-1mm}
\end{abstract}
\maketitle

\section{Introduction}

The zone method approach to the fermion sign problem is based on the
observation that in many finite temperature simulations fermion permutations
are short ranged \cite{lee}. 
This can be true even for systems with massless modes and
long distance correlations, as we demonstrate with an explicit example.  
In this proceedings article we derive an estimate for the finite temperature 
fermion wandering length and discuss the main features of the zone 
method.

\section{Worldlines}

We begin with a brief review of the worldline formalism \cite{hirsch}. \ We
introduce ideas in one spatial dimension before moving on to
higher dimensions. \ Let us consider a system with one species of fermion on
a periodic chain with $L$ sites, where $L$ is even. \ Aside from an additive
constant, the general Hamiltonian can be written as 
\begin{equation}
H=-h\sum_{i}\left[ a_{i+1}^{\dagger }a_{i}+a_{i}^{\dagger }a_{i+1}\right]
+\sum_{i}c_{i}a_{i}^{\dagger }a_{i}.
\end{equation}
Following \cite{hirsch} we break the Hamiltonian into two parts, $H_{\text{e}%
}$ and $H_{\text{o}}$, 
\begin{equation}
H_{\text{e/o}}=\sum_{i\text{ even/odd}}\left[ 
\begin{array}{c}
-ha_{i+1}^{\dagger }a_{i}-ha_{i}^{\dagger }a_{i+1} \\ 
+%
{\textstyle{c_{i} \over 2}}%
a_{i}^{\dagger }a_{i}+%
{\textstyle{c_{i+1} \over 2}}%
a_{i+1}^{\dagger }a_{i+1}
\end{array}
\right] .  \nonumber
\end{equation}
We note that $H=H_{\text{e}}+H_{\text{o}}$.  For large $N$, we can write 
\begin{equation}
\exp (-\beta H) \approx (S_{\text{o}}S_{\text{%
e}})^{N},  \label{exp}
\end{equation}
where 
\begin{equation}
S_{\text{e/o}}=\exp (-%
{\textstyle{\beta  \over N}}%
H_{\text{e/o}}).
\end{equation}
Inserting a complete set of states at each step, we can write $Tr\left[ \exp
(-\beta H)\right] $ as 
\begin{equation}
\sum_{z_{0},...,z_{2N-1}}\left\langle z_{0}\right| S_{\text{o}}\left|
z_{2N-1}\right\rangle ...\left\langle z_{1}\right| S_{\text{e}}\left|
z_{0}\right\rangle .  \label{b}
\end{equation}

The worldline trajectory of each of the fermions can now be traced from Euclidean time $t=0$ to time $t=\beta $.  Since we are computing a thermal trace, the worldlines from $t=0$ to $t=\beta $
define a permutation of identical fermions. \ Even permutations carry a
fermion sign of $+1$ while odd permutations carry sign $-1$.  The
generalization to higher dimensions is straightforward.  In two dimensions,
for example, $\exp (-\beta H)$ takes the form 
\begin{equation}
\left. 
\begin{array}{c}
\left[ \exp \left[ -%
{\textstyle{\beta  \over N}}%
(H_{\text{e}}^{x}+H_{\text{o}}^{x}+H_{\text{e}}^{y}+H_{\text{o}}^{y})\right] 
\right] ^{N} \\ 
\qquad \qquad \qquad \approx (S_{\text{o}}^{y}S_{\text{e}}^{y}S_{\text{o}%
}^{x}S_{\text{e}}^{x})^{N}.
\end{array}
\right.   \label{2d}
\end{equation}

\section{Wandering length}

In one spatial dimension, the Pauli exclusion principle inhibits fermion
permutations except in cases where the fermions wrap around the lattice
boundary. \ For the remainder of our discussion, therefore, we consider
systems with two or more dimensions. \ The first question we address is how
far fermion worldlines can wander from start time $t=0$ to end time $t=\beta$%
. \ We can put an upper bound on this wandering distance by considering the
special case with no on-site potential and only nearest neighbor hopping.

Let us consider motion in the $x$-direction. \ For each factor of $S_{\text{o%
}}^{x}S_{\text{e}}^{x}$ in (\ref{2d}) a given fermion may remain at the same 
$x$ value, move one lattice space to the left, or move one lattice space to
the right. \ If $h$ is the hopping parameter, then for large $N$ the
relative weights for these possibilities are approximately $1$ for remaining
at the same $x$ value, $\beta hN^{-1}$ for one move to the left, and $\beta
hN^{-1}$ for one move to the right. \ In (\ref{2d}) we see that there are $N$
factors of $S_{\text{o}}^{x}S_{\text{e}}^{x}$. \ Therefore for a typical
worldline configuration at low filling fraction, $f,$ we expect $\sim\beta h$
hops to the left and $\sim\beta h$ hops to the right. \ For non-negligible $%
f $ some of the hops are forbidden by the exclusion principle. \ Assuming
random filling we expect $\sim(1-f)\beta h$ hops to the left and $%
\sim(1-f)\beta h$ hops to the right.

The net displacement is equivalent to a random walk with $2(1-f)\beta h$
steps. \ The expected wandering length, $\Delta,$ is therefore given by 
\begin{equation}
\Delta=\sqrt{2(1-f)\beta h}.  \label{wandering}
\end{equation}

In cases with on-site potentials, fermion hopping is dampened by differences
in potential energy. \ Hence the estimate ($\ref{wandering}$) serves as an
upper bound for the general case. \ We have checked the upper bound
numerically using simulation data generated by many different lattice
Hamiltonians with and without on-site potentials.

\section{Permutation Zone Method}

Let $W$ be the logarithm of the partition function, 
\begin{equation}
W=\log \left\{ Tr\left[ \exp (-\beta H)\right] \right\} .
\end{equation}
Let us partition the spatial lattice, $\Gamma $, into zones $%
Z_{1},Z_{2},...,Z_{k}$ such that the spatial dimensions of each zone are much
greater than $\Delta $. \ For notational convenience we define $%
Z_{0}=\emptyset $. \ For any $R\subset \Gamma $, let $W_{R}$ be the
logarithm of a restricted partition function that includes only worldline
configurations such that any worldline starting at $t=0$ outside of $R$
returns to the same point at $t=\beta $. \ In other words there are no
permutations for worldlines starting outside of $R$. \ We note that $%
W_{\Gamma }=W,$ and $W_{\emptyset }$ is the logarithm of the restricted
partition function with no worldline permutations at all. \ Since the zones
are much larger than the length scale $\Delta $, the worldline permutations
in one zone has little or no effect on the worldline permutations in another
zones. \ Therefore
\begin{equation}
W_{Z_{0}\cup... Z_{j}}-W_{Z_{0}\cup... Z_{j-1}}\approx
W_{Z_{j}}-W_{\emptyset }.
\end{equation}
Using a telescoping series, we obtain 
\begin{equation}
W_{\Gamma}-W_{\emptyset}  = \sum_{j=1,...,k}(W_{Z_{0}\cup... 
Z_{j}}-W_{Z_{0}\cup... Z_{j-1}}),
\end{equation}

\begin{equation}
W_{\Gamma }-W_{\emptyset } 
\approx \sum_{j=1,...,k}(W_{Z_{j}}-W_{\emptyset }). \label{est}
\end{equation}
For translationally invariant systems tiled with congruent zones we find 
\begin{equation}
W=W_{\Gamma }\approx W_{\emptyset }+%
{\textstyle{\left| \Gamma \right|  \over \left| Z_{1}\right| }}%
(W_{Z_{1}}-W_{\emptyset }),  \label{tile}
\end{equation}
where $\left| \Gamma \right| /\left| Z_{1}\right| \equiv k$, the number of
zones. \ For general zone shapes one can imagine partitioning the zones
themselves into smaller congruent tiles. \ Therefore the result (\ref{est})
should hold for large arbitrarily shaped zones. \ For this case we take $%
\left| \Gamma \right| $ to be the number of nearest neighbor bonds in the
entire lattice and $\left| Z_{1}\right| $ to be the number of nearest
neighbor bonds in the zone. \ We will refer to $\left| Z_{1}\right| $ as the
zone size of $Z_{1}$. \ This is just one choice for zone extrapolation. \ A
more precise and complicated scheme could be devised which takes into
account the circumscribed volume, number of included lattice points, and
other geometric quantities.

As an example of the permutation zone method, we compute the average energy $%
\left\langle E\right\rangle h^{-1}$ for a free fermion Hamiltonian with only
hopping on an $8\times 8$ lattice. \ We set the number of time steps $N=8$
and consider values $\beta h=1.0$, $1.5$, and $2.0$. \ The corresponding
values for $\Delta $ are $1.0$, $1.2$, and $1.4$ respectively. \ The Monte
Carlo updates were performed using our own version of the single-cluster
loop algorithm \cite{evertz}.
In Figure 1 we show data for rectangular zones with side dimensions $0\times 0$%
, $1\times 1$, $2\times 1$, $2\times 2$, $3\times 2$, $3\times 3$, ..., $%
6\times 6$. \ We also show a least-squares fit (not including the smallest
zones $0\times 0$ and $1\times 1$) assuming linear dependence on zone size
as predicted in (\ref{est}). \ We find agreement at the $1\%$ level or
better when compared with the exact answers shown on the far right, which
were computed using momentum-space decomposition.

While the physics of the free hopping Hamiltonian is trivial, the
computational problems are in fact maximally difficult. \ The severity of
the sign problem can be measured in terms of the average sign, $<$Sign$>$,
for contributions to the partition function. \ For $\beta h=1.0$, $<$Sign$%
>\sim0.005;$ for $\beta h=1.5$, $<$Sign$>$ $\sim10^{-6};$ and for $\beta
h=2.0$, 
\mbox{$<$}%
Sign%
\mbox{$>$}%
$\sim10^{-9}$. \ Direct calculation using position-space Monte Carlo is
impossible by several orders of magnitude for $\beta h\geq1.5$.

\begin{figure}[t]
\epsfig{file=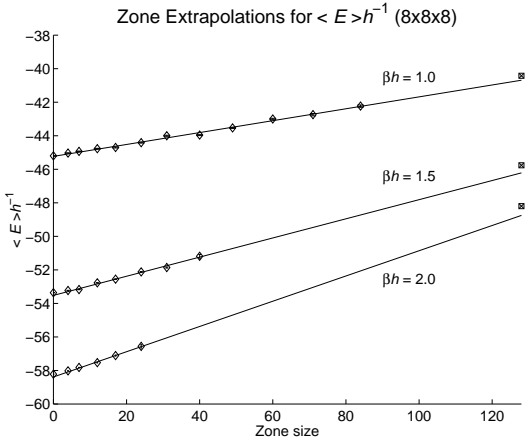,width=7cm}
\caption{Average energy at $\protect\beta h=1.0,1.5,2.0$ for the free
fermion Hamiltonian on an $8\times 8$ lattice$.$}
\end{figure}

\section{Summary}

We have reviewed the zone method approach to the fermion sign problem.  We
have demonstrated that the exchange of identical fermions is short-ranged
and has a maximum range of $\sqrt{2(1-f)\beta h}$ lattice sites, where $%
\beta $ is the inverse temperature, $h$ is the hopping parameter, and $f$ is
the filling fraction.  We have introduced the notion of permutation zones,
special regions of the lattice where identical fermions may interchange and
outside of which they may not.  Using successively larger permutation
zones, one can extrapolate to obtain thermodynamic observables.  Applications 
of the zone method to chiral symmetry breaking in (2+1) dimensional QED is 
discussed in another article in these proceedings \cite{maris}.

\end{document}